\documentclass[letterpaper,twocolumn,10pt]{article}
\usepackage{usenix2024_SOUPS}

\usepackage{tikz}
\usetikzlibrary{arrows.meta, positioning}
\usepackage{amsmath}

\usepackage{filecontents}
\usepackage{array}
\usepackage{booktabs}
\usepackage{colortbl}
\usepackage{xcolor}
\usepackage{xurl}

\usepackage{longtable}
\usepackage{tabularx}
\usepackage{ragged2e}
\usepackage{enumitem}
\usepackage{cuted}     
\usepackage{capt-of}

\newcommand{\highlight}[1]{\vspace{.03in}\noindent{\bf #1}.}

\newcommand{\plainauthor}{Francis Hahn, Mamoon Mohd, Alex Bardas, Michael Collins, Daniel Lende, Xinming Ou, S. Raj Rajagopalan}
\begin{document}

\date{}

\title{\Large \bf A Sociotechnical, Practitioner-Centered Approach to Technology Adoption in Cybersecurity Operations: An LLM Case}

\author{
\begin{tabular}[t]{ccc}
{\rm Francis Hahn\thanks{Francis Hahn and  Mohd Mamoon contributed equally and share first authorship.}} & {\rm Mohd Mamoon\footnotemark[1]} & {\rm Alexandru G. Bardas} \\
\textit{\small University of South Florida} & \textit{\small University of Kansas} & \textit{\small University of Kansas} \\
[0.4em]
{\rm Michael Collins} & {\rm Daniel Lende} & {\rm Xinming Ou} \\
\textit{\small Univ. of Southern California - ISI} & \textit{\small University of South Florida} & \textit{\small University of South Florida} \\
[0.4em]
\multicolumn{3}{c}{{\rm S. Raj Rajagopalan}} \\
\multicolumn{3}{c}{\textit{\small Resideo Technologies}}
\end{tabular}
}
\maketitle

\begin{abstract}

Technology for security operations centers (SOCs) has a storied history 
of slow adoption due to concerns about trust and reliability. These concerns 
are amplified with artificial intelligence, particularly large language models (LLMs), 
which exhibit issues such as hallucinations and inconsistent outputs. 
To assess whether LLM-based tools can improve SOC efficiency, 
we embedded two PhD researchers within a multinational company SOC for six months of 
ethnographic fieldwork. We identified recurring challenges, such as repetitive tasks, 
fragmented/unclear data, and tooling bottlenecks, and collaborated directly 
with practitioners to develop LLM companion tools 
aligned with their operational needs. 

Iterative refinement reduced workflow disruption and improved interpretability, 
leading from skepticism to sustained adoption. Ethnographic analysis indicates that 
this shift was enabled by our sociotechnical co-creation process consistent with 
Nonaka's SECI model. 
This framework explains the common challenges
in traditional SOC technology adoption, including workflow misalignment, 
rigidity against evolving threats and internal requirements, 
and stagnation over time. 
Our findings show that the co-creation approach can overcome
these old barriers and create a new paradigm for creating usable technology
for cybersecurity operations.

\end{abstract}

\section{Introduction}
Security Operations Centers (SOCs) are foundational to modern 
information security across industry, academia, and government. 
They serve as the operational hub for monitoring network activity, 
defending against cyber threats, and supporting regulatory 
compliance ~\cite{Sundramurthy2015SOUPS,sundaramurthy2017humans}. 
Effective SOC operations rely on the coordinated interaction of people, 
processes, and technology, working together to enable timely detection, 
investigation, and response to security incidents~\cite{zimmerman2014ten}.

Traditionally, a key challenge for SOC analysts has been determining 
whether an alert or finding is a false positive or a true 
positive~\cite{alahmadi202299,johnson2013don,smith2015questions}. 
This often requires navigating vast, distributed data across 
multiple platforms and reaching conclusions under strict time 
constraints~\cite{tariq2025alert,kersten2025field}. 
In the context of this high-stress work environment, introduction of
any new technology is often viewed with great suspicion. A new tool
for SOC could result in disruption of the existing workflow with ultimately 
little or no real efficiency gain~\cite{sundaramurthy2017humans}. 

There has been increasing interest in
applying generative AI (GenAI) such as large language models (LLMs) in
cybersecurity~\cite{silva2024llmcybersurvey,xu2024llm4securityslr}.
In fact, some leading companies are already pushing out products
and services that purportedly incorporate 
GenAI~\cite{Freitas2025ACM,mscopia_rct2024,rapid7aiengine2024,reliaquestaiagent2024}.
Practitioners' reception to such technology has been cautious~\cite{Roch2024Usenix},
due to well-known problems such as hallucinations and incorrect/inconsistent
answers.

Despite long-standing efforts to improve SOC effectiveness and reduce 
burnout~\cite{thimmaraju2025human} through improved processes, tools, 
and training, research consistently identifies a persistent gap between 
academic work and operational SOC practice (e.g.,~\cite{Kokulu2019CCS},~\cite{Catal2022ACM}). 
Ethnographic studies show that tools designed without accounting for 
analysts' experiential and iterative reasoning often fail in practice, 
leading to limited adoption or abandonment~\cite{Werlinger2009IJHCS}. 
This reflects a recurring mismatch between practitioners' needs and 
researchers' assumptions.

We draw on Nonaka's SECI model of knowledge conversion 
(Socialization, Externalization, Combination, Internalization)~\cite{Nonaka94} 
and its adaptation to cybersecurity fieldwork through Tool-Oriented SECI~\cite{sundramurthy2014IEEE}.
We examine how LLM-based companion systems fit within and extend this 
knowledge conversion cycle. In particular, these systems generate new 
explicit artifacts that must undergo trust calibration before analysts 
internalize and incorporate them in workflow.

As part of this effort, we embedded two 
computer science PhD students in a major commercial organization. 
Through participant observation and sustained fieldwork, we identified where explicit and tacit knowledge resides and co-developed LLM-based companion tools with SOC analysts to meet operational needs. 
Using a sociotechnical lens, we analyzed adoption patterns and
iterative refinement driven by practitioner feedback, resulting in a
AI companion platform that supported multiple security tools for the
analysts. This iterative engagement mirrored and extended the SECI
cycle.  Embedded participation surfaced tacit workflow expectations
(Socialization); reflection through discussions and interviews
highlighted pain points and challenges (Externalization); tool
development incorporated these insights into usable tools
(Combination); and deployment enabled trust-calibrated internalization
within operational routines (Internalization).
Over the six-month engagement, our student researchers participated in
on-boarding and training, attended meetings, and collaborated closely
with SOC analysts and managers. This sustained presence enabled us to
observe workflows, tool usage, communication patterns, and breakdowns
from an insider perspective. Sustained day-to-day
interaction~\cite{Lave_Wenger_1991} fostered trust, and provided
access to candid operational insights typically unavailable to
external researchers.
By applying a sociotechnical perspective to the challenges faced by 
cybersecurity professionals, we extend prior work through the explicit 
use of co-creation as a structured method for tool development and integration. 
Accordingly, the \textbf{central research theme} of this work can be summarized  
as a practitioner centered, sociotechnical investigation of how LLM based 
tools can be co-created, trusted, and embedded into cybersecurity operations 
to improve workflow efficiency while preserving privacy, security, and 
operational reliability. 

Through participant observation, fieldwork, and co-creation, we show that a 
sociotechnical approach grounded in direct collaboration with cybersecurity 
professionals improves both the adoption and operational impact of LLMs.
Our findings demonstrate how social observation and technical development are 
tightly integrated. 
Our contributions can be summarized as follows: 
\vspace{-.05in}
\begin{itemize}[leftmargin=*,topsep=2pt,itemsep=2pt,parsep=0pt]

    \item A structured co-creation framework for developing LLM-based tools with cybersecurity practitioners in live operational environments. The approach integrates fieldwork, participant observation, and tool development to improve operational relevance and impact.

   \item Illustrate how thematic analysis supports the systematic identification of workflow pain points, operational challenges, and analyst reception to the process.

    \item Provide insights into adoption and long-term sustainability through a staged integration strategy that incrementally built trust and operational alignment within an industry SOC. 
     \item Extend Tool-Oriented SECI by identifying generative recombination and 
    trust-calibrated internalization as defining characteristics of LLM-mediated 
    knowledge conversion.

\end{itemize}


\section{Related Work}
This section reviews prior work across three areas: embedded sociotechnical studies of SOC practice, AI and LLM-based tools for security operations, and analyst-centered work on trust, explainability, and workflow integration.

\highlight{Embedded Sociotechnical Studies in SOCs}
Prior research at the intersection of cybersecurity, sociotechnical systems, and researcher-practitioner collaboration has consistently documented gaps between academic security research and operational practice~\cite{Haney2024SOUPS}. Ethnographic studies show that SOC work is practice-driven, socially organized, and shaped by informal coordination as much as by technical tooling~\cite{Sundramurthy2016SOUPS}. Sundaramurthy~et~al.~\cite{Sundramurthy2016SOUPS}, drawing on a 3.5-year multi-site anthropological study, model SOC work as an activity-theoretic system shaped by tensions among tools, procedures, and organizational expectations. Their findings emphasize that meaningful innovation requires embedded researchers who understand analysts’ tacit knowledge and operational constraints.

Previous work also identifies sociotechnical barriers to AI adoption in cybersecurity. Roch et al.~\cite{Roch2024Usenix} show that although experts are willing to use AI for repetitive and data-intensive tasks, adoption is limited by concerns about trust, transparency, determinism, and data protection. Analysts favor gradual trust-building through low-risk task delegation and calibrated autonomy. Similarly, Fung et al.~\cite{fung2025adopting} emphasize that trust, workflow alignment, and secure, closed deployments are critical for successful integration. These studies highlight that effective AI adoption in security contexts depends on trust calibration and organizational fit.

Beyond SOCs, Tuladhar et al's~\cite{Tuladhar2021SOUPS} insider ethnography illustrates how secure practices become organizational norms only when learned in context and reinforced through hands-on collaboration with security advocates. Vielberth et al.’s~\cite{Vielberth2020SecurityOC} systematic review similarly finds that SOC research remains fragmented, often relying on interviews or isolated case studies and lacking holistic sociotechnical frameworks. Greig et al.~\cite{Greig2015CIS} further document persistent misalignments between formal security policies and everyday practice.

Collectively, this literature establishes that organizational
cybersecurity policies are enacted through people’s workflows, stories,
and peer norms rather than through tools alone. However, while these
studies richly characterize SOC work, they do not examine how new AI
tools are shaped through co-creation or integrated into live workflows
of cybersecurity professionals over time. As a result, we lack
empirical understanding of how sociotechnical dynamics influence the
adoption of LLM companions in operational security settings.

\highlight{AI/LLM Tools for Security Operations}
Recent research has explored how AI, and more recently LLMs, can
support tasks across the security operations lifecycle. Prior work has
largely focused on applying machine learning to detection-centric
problems such as alert ranking, anomaly detection, and log analysis,
with mixed success in operational settings. Srinivas et
al.~\cite{Srinivas2025JCSP} presents a comprehensive taxonomy of AI
applications in SOCs, cataloging use cases ranging from log
summarization and alert triage to vulnerability management. Despite
this breadth, they highlight persistent challenges including
interpretability, robustness, hallucinations, and legacy system
integration that continue to limit operational fit.
Binbeshr~et~al.~\cite{Binbeshr2025CongitiveSOCs} similarly note that while
ML-driven approaches dominate the literature and promise improved
detection accuracy, but their deployment remains constrained by data
quality, explainability, and trust.

Empirical studies consistently show that practitioners view AI systems
as augmentative rather than autonomous. Mink~et~al.~\cite{Mink2023Oakland}
find that analysts prefer ML tools that
complement existing rule-based workflows and provide contextual
explanations that support verification, rather than opaque
predictions. In the LLM context, Singh et al.~\cite{singh2025Archive}
reported from a 10-month longitudinal study of 45 analysts that GPT-4
was primarily used as an on-demand sensemaking aid for interpreting
commands, scripts, and telemetry rather than as a decision
authority. These findings suggest that LLMs are most valuable when
embedded into analysts’ reasoning processes, rather than positioned as
replacements for expert judgment.

Recent systems demonstrate the growing interest in LLM-based security tooling. Kramer et al.~\cite{kramer2025integrating} examined the use of LLMs for incident response summarization. Through controlled experiments with 18 analysts and 50 real-world incidents, they show that fully autonomous LLM-generated summaries frequently omit critical details or introduce factual inaccuracies, but that collaborative, analyst-in-the-loop use can improve readability and consistency while reducing effort. Importantly, their work evaluates LLMs primarily as post-hoc summarization aids, and focuses on controlled experimental settings rather than live, evolving SOC workflows.

Other work explores LLMs as task-oriented security assistants. Deng et
al.’s PentestGPT~\cite{pentestgpt} demonstrates LLM driven
orchestration of penetration testing workflows by chaining tools and
reasoning steps, while Shashwat et al.~\cite{shashwat2024preliminarystudyusinglarge}
explore early applications for security analysis and triage. Although
these systems highlight LLM flexibility as reasoning and orchestration
layers, they are largely evaluated in isolated or simulated settings,
with limited focus on organizational constraints, analyst trust, or
sustained adoption.

Microsoft’s Copilot for Security further illustrates the scalability of
LLM-based approaches, reporting improvements in aggregate triage and
remediation metrics~\cite{Freitas2025ACM}. However, such evaluations
largely emphasize system-level performance and leave unanswered
questions about how these tools reshape individual analyst workflows,
decision-making practices, and accountability in day-to-day
operations.

Across this literature, LLMs are shown to be promising sensemaking
and productivity aids. Yet none of these studies attempted to investigate
LLMs' adoption in real operational environments through zero-proximity
observations and active researcher interventions.

\highlight{Analysts and Sociotechnical Factors}
A body of research emphasizes the importance of analyst-centered
design, trust, and workflow integration in security tooling.
Rastogi~et~al.~\cite{Rastogi2025Archive} show that generic explainability
techniques often fail to meet analysts’ needs in high-pressure SOC
contexts, where time constraints and alert volume demand concise,
actionable explanations. Nyre-Yu~et~al.~\cite{NyreYu2022USEC} report
that a deployed explainable-AI tool saw limited use and did not
improve analyst accuracy, underscoring the importance of aligning
system outputs with users’ roles, responsibilities, and trust
thresholds.

These findings echo broader human-AI collaboration research, which emphasizes maintaining user authority, calibrating trust, and embedding AI outputs into established practices rather than displacing them. However, prior work typically examines trust and explainability as static design properties, rather than as outcomes negotiated through sustained use and organizational context. 

\highlight{Our Work} 
In contrast to prior work, our study integrates embedded ethnography
with the co-creation and in-situ deployment of LLM-based companion
tools within a live SOC. This design enables us to examine how
analysts perform their work, and how LLM tools are iteratively
developed, their outputs negotiated, validated, and internalized
within operational workflows over time. By tracing this integration
process, we address a gap in SOC and AI-for-security research, which
often evaluates systems outside the contexts in which they must
ultimately function.


\section{Methodology}
This study examines the design and integration of LLM-based companion tools for domain-specific security tasks within a live SOC. Using a sociotechnical approach, researchers were embedded within an industry SOC to understand workflows, norms, and operational constraints through fieldwork and enculturation~\cite{Lende2012MIT}. Co-creation with practitioners provided context-rich feedback that guided iterative development.

\begin{figure}[t]
  \centering
  \includegraphics[width=1\columnwidth,trim=5 5 5 5,clip]{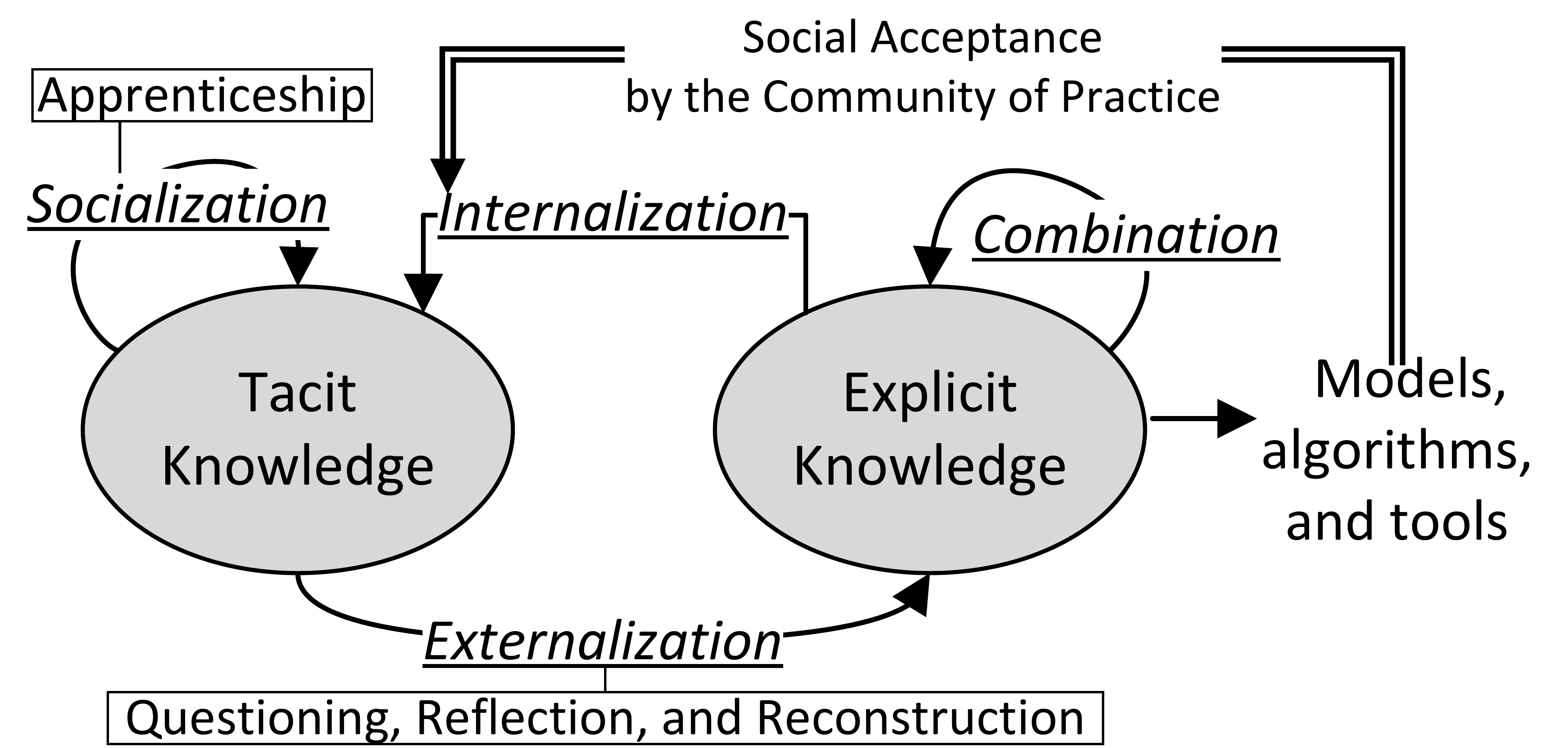}
  \vspace{-.1in}
  \caption{Tool-Oriented SECI Model (based on~\cite{sundramurthy2014IEEE}) -- SECI adaptation for cybersecurity fieldwork where ethnographic insight is translated into operational tools that embed explicated tacit knowledge back into analyst workflows.}
  \label{figure:SECI-model}
\end{figure}

\highlight{SECI-Informed Research Design}
Our methodological approach was informed by Nonaka’s SECI
model~\cite{Nonaka94,nonaka2009perspective} of knowledge conversion and its
tool-oriented adaptation for cybersecurity
fieldwork~\cite{sundramurthy2014IEEE}, see
Figure~\ref{figure:SECI-model}. Rather than treating ethnographic
observation and tool development as separate activities, we structured
our field engagement as an iterative cycle of knowledge conversion
mediated by tool building.  Overall, (i) fieldworkers acquired tacit
operational knowledge through embedded participation
\textit{(Socialization: tacit$\rightarrow$tacit)}, (ii) surfaced and
articulated implicit pain points and challenges
\textit{(Externalization: tacit$\rightarrow$explicit)}, (iii)
formalized these challenges into usable tools \textit{(Combination:
explicit$\rightarrow$explicit)}, and (iv) evaluated how these tools
shaped practice through deployment and iterative adoption
\textit{(Internalization: explicit$\rightarrow$tacit)}.

\highlight{Ethnographic Approach}
Participant observation, fieldwork, and co-creative engagement were central to our ethnographic approach. Through close engagement, we examined how SOC analysts perform their work, integrated tools in their workflows, and coordinate within their environment. 
By participating in peripheral tasks~\cite{Lave_Wenger_1991} and encountering situated challenges~\cite{suchman1984plans}, we gained firsthand insight into operational constraints. These experiences, combined with interviews and informal discussions, provided a context-rich understanding of the daily challenges analysts face in security operations. 

Beyond observation, fieldworkers sought participation in analyst tasks. This included requesting access to relevant databases, vendor platforms, and APIs, and submitting proposals to deploy LLM-based tools within the organization’s infrastructure. This provided hands-on exposure to the technical and institutional constraints shaping daily operations.

\begin{figure*}[t]
  \centering
  \includegraphics[width=\textwidth]{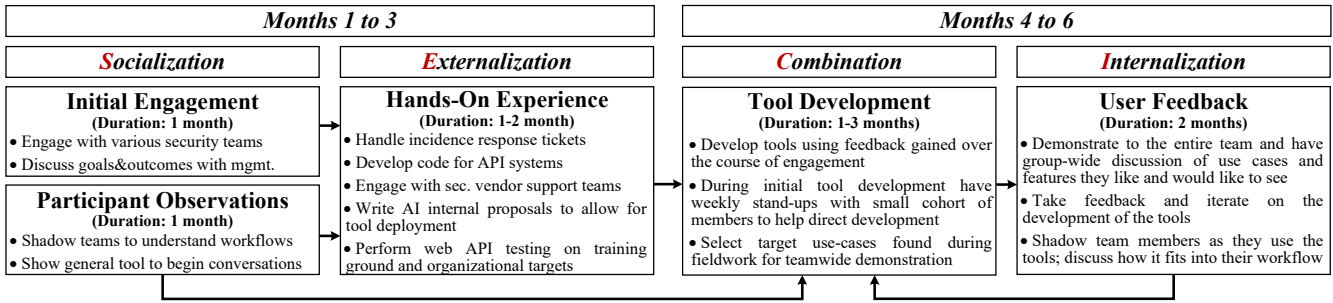}
  \vspace{-.2in}
  \caption{Feedback Loop -- Six-month embedded fieldwork and co-creation timeline showing how ethnographic immersion, hands-on SOC work, and iterative user feedback formed a continuous feedback loop for tool design and refinement}
  \label{fig:feedbackLoopModel}
\end{figure*}

 \highlight{Iterative Feedback Loop} A key component of our ethnographic methodology was a structured feedback-loop process (Figure~\ref{fig:feedbackLoopModel}) that incorporated insights gathered during demonstrations, shadowing, and live tool use. The loop enabled continuous refinement of both the tools and fieldworker's understanding of analyst work. During each iteration, they adapted and tuned LLM configurations to specific use cases and conducted comparative evaluations across models to determine suitability prior to deployment. 


\highlight{SECI Phase 1: Socialization (Tacit$\rightarrow$Tacit)}
During the first month of engagement, the fieldworkers sought to develop an external perspective on the security team while gradually building trust through sustained social interaction. Because the environment was fully remote, traditional forms of participant observation, such as informal conversations or spontaneous workplace interactions, were absent. Instead, observation was conducted through Microsoft Teams meetings, screen sharing, and text-based communication. Organizational artifacts, including hierarchy maps, were used to understand reporting structures, identify stakeholders, and coordinate engagement across geographically distributed teams.

\highlight{The Internal Advocate (IA)}
The engagement began by connecting with members across the organization, facilitated by an internal advocate (IA) who provided legitimacy to the fieldworker’s presence and helped expand their access within the organization~\cite{engestrom2017studies, van2011tales}. Early in the embedding, the researchers relied on the IA’s organizational knowledge and hierarchical position to navigate internal dynamics, communicate goals with management and team members, and identify key pain points and high-priority issues.

\highlight{SECI Phase 2: Externalization (Tacit$\rightarrow$Explicit)}
To externalize tacit understanding acquired through the embedding, the
fieldworkers conducted structured reflection and reconstruction of
analyst workflows. This reflexive step was supported by (i) workflow
walkthroughs (step-by-step demonstrations across vendor platforms and
internal tools), (ii) debrief discussions following observations and
demo sessions, and (iii) targeted questioning to surface day-to-day
challenges and pain points. Most important, the fieldworker participated
in the same tasks as the other analysts. This hands-on experience is
necessary to the reflection and reconstruction process. Externalization outputs included written
workflow narratives, reconstructed investigative models (e.g.,
query-building sequences, cross-system data access patterns), and
explicit articulation of “pain points” such as repetitive log
correlation, and manual cross-referencing across systems.

\highlight{SECI Phase 3: Combination (Explicit$\rightarrow$Explicit)}
In our fieldwork, the combination phase manifests as the creation
of AI companions that incorporate the externalized tacit knowledge.
Drawing on insights from participant observation and fieldwork, the
fieldworkers identified several LLM-supported tools to address
tedious, repetitive tasks involving widely distributed or
difficult-to-interpret data within the SOC and pentesting teams. These
tools targeted Root Cause Analysis (RCA), Asset Discovery, and Query
Building. Development progressed iteratively, with incremental
demonstrations during team stand-ups to share updates, gather
feedback, and discuss security implications and future design
directions.

\highlight{Collecting Data}
The systematic recombining of explicit artifacts was enabled by
document collection from the SOC analyst team at different stages of
tooling co-creation. In the initial stages the collected data existed
as static resources, including RCA playbooks and query examples from
the security vendor. After evaluating initial tool performance, data
collected was supplemented with custom examples derived from
fieldwork insights. In the final phase, analysts contributed
real-world examples directly. Combination consisted of systematically
recombining explicit artifacts collected from the field (playbooks,
internal documents) with explicit reconstructions produced during
Externalization (prompt templates, query patterns, RCA reports, and so on).

\highlight{SECI Phase 4: Internalization (Explicit$\rightarrow$Tacit)}
The internalization process is realized through deployment and itearative
co-adaptation of the developed AI companions. 
Our co-creation approach followed an iterative design process grounded
in continuous engagement with security team members, drawing on their
tacit knowledge and making it explicit through discussion and
collaboration~\cite{ehn1988work}. Regular testing and demo sessions
were used to evaluate usefulness, identify limitations, and refine
features for integration into daily workflows. This deliberate
co-creative process ensured that the resulting tools reflected both
the fieldworker’s insights and the practical expertise of
cybersecurity professionals.

In this study, co-creation is a process in which researchers and
cybersecurity professionals jointly develop, evaluate, and refine
LLM-based tools while simultaneously synthesizing knowledge through
close, ongoing collaboration~\cite{Anderson2016MJA,
Greenhalgh2016Millbank}.


\highlight{Participant Recruitment} Participants included 18 SOC analysts, including CISO and the SOC director, and members of adjacent security teams (e.g., vulnerability management and pentesting). Sampling was purposive and within the constraints of organizational access, prioritizing roles directly involved in alert handling, query construction, and tool use.

\highlight{Thematic Analysis}
Both fieldworkers independently coded field notes, meeting transcripts, feedback, and observational artifacts using an inductively developed codebook that was iteratively refined throughout the analysis. They iteratively applied the codebook until no new codes and themes emerged~\cite{Panahi2024Usenix}. The
codebook is included in the appendix.

\section{LLM Opportunity Identified through Ethnographic Study}
By working alongside the
organization’s security team, the PhD researchers learned
operational norms, practices, and constraints necessary for developing
impactful AI- and LLM-based tools. At the outset, the organization’s
exposure to AI was largely limited to vendor-provided modules. To
acquire the situated knowledge required for effective tool design, we
positioned fieldworkers as
learners in close collaboration with experienced analysts.
Through the ethnographic study, we identified a number of
pain points in the operation that offer opportunities for
LLM adoption.





\begin{figure*}[t]
  \centering
  \includegraphics[
      width=\textwidth,
      trim=0.5cm 7.5cm 0.5cm 7.5cm,  
      clip
  ]{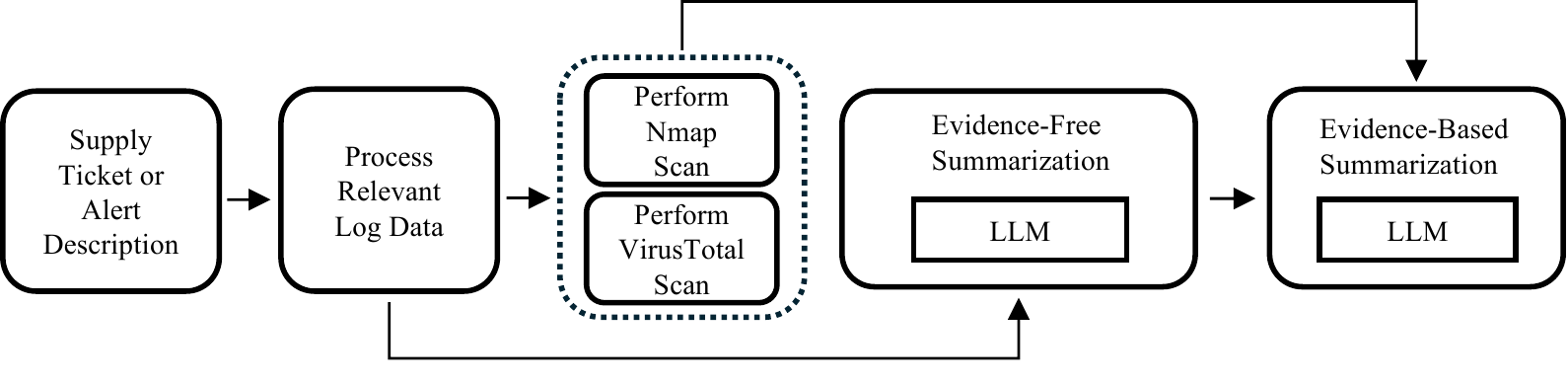}
  \vspace{-.25in}
  \caption{Demonstration SOC Tool Workflow -- End-to-end pipeline showing how alert inputs are processed, enriched with external scans, and summarized by an LLM with and without supporting evidence.}
  \label{fig:feedbackLoop}
\end{figure*}
\subsection{Tooling Gaps}
Across discussions, a consistent theme emerged: limitations and bottlenecks caused by the security vendor’s platform, siloed data sources, and ad-hoc task assignments. Many tooling constraints were rooted in organizational policies that dictated how cybersecurity professionals were required to perform their work. 
Throughout the embedded study, the fieldworkers identified several recurring workflow gaps, including:
\begin{enumerate}[topsep=2pt,itemsep=2pt,parsep=0pt]
    \item  Developing automation workflows that leverage organizational data not accessible to the vendor’s platform.
    \item  Handling large quantities of data, logs or source code, that is often obfuscated or unintelligible.
    \item  Building complex query statements for the platform.
    \item  Handling distributed data across multiple platforms.
    \item  Maintaining the security vendor's visibility of assets. 
\end{enumerate}
\highlight{Root Cause Analysis (RCA)}
RCA is an event that takes place between the SOC, legal, management, and any other team which have been the victim of an incident which resulted in the loss, alteration, or unintended addition of data to the organization's data stores. This has been marked as a time critical event which involves participants with a wide-range of technical background, perspectives, objectives, and involvement in the incident. 

Through our discussions with both the legal team and the SOC we identified RCA as a strong use case where LLMs could assist. The idea to use LLMs to assist with RCA was shared by everyone due to the time critical nature of the event and the need to summarize the data being analyzed for the various perspectives and objectives. Due to the sensitivity of the data, it was often not something the security platform could assist with directly, but would be involved for collecting or searching for data. The RCA tool was designed to assist with gaps \textbf{(1)}, \textbf{(2)}, and \textbf{(4)}.

\highlight{Query Builder}
During the fieldwork the embedded
researchers noticed a need for assistance with building query
statements to search for log and event information when performing
incident response for the alerts they were assigned using the
security vendor's security information and event management system
(SIEM). During the time of the fieldwork, the security vendor
released an AI-powered query builder for assisting
with performing query statement construction.
However, due to limited visibility into critical organizational event
logs and the inability to supply this information, the query builder
was only able to assist in constructing query statements similar to
examples found in online documentation. Discussions with analysts
revealed that their experiences with the tool were similar. This led
the fieldworkers to consider developing an internal query-building
tool capable of leveraging user-supplied examples and storing them for
future use, creating a living database that would expand the available
knowledge base over time. The query builder was initially designed to
address gaps \textbf{(1)} and \textbf{(3)}, and during the iterative
refinement phase, it was extended to address gap \textbf{(2)}.

\highlight{Asset Discovery}
Following the demonstration of the model SOC tool, discussions with a vulnerability management analyst surfaced interest in the system’s data processing pipeline, particularly how structured data from Excel files was parsed, aggregated, and prepared for LLM-based summarization. These discussions prompted closer examination of the organization’s asset discovery workflow.

The workflow involved polling the security vendor’s API to retrieve agent status information, providing visibility into organizational assets. Retrieved data was then correlated with records from internal databases (e.g., Azure and ServiceOne) and consolidated into remediation emails sent to asset owners to address reporting inconsistencies. To develop a grounded understanding of this process, fieldworkers transitioned from observation to direct task execution, including implementing API interactions and working with the relevant databases. This hands-on engagement enabled end-to-end understanding of the workflow and informed subsequent tool design decisions.

Participant observation and fieldwork revealed several recurring challenges. Data was highly distributed across multiple systems. It was siloed and often inconsistent. Analysts needed to perform recurring API calls to an application at fixed intervals. They also had to aggregate and summarize the collected information in order to produce routine reports and track remediation actions for identified issues. 
These observations led to discussions about designing and developing an asset discovery tool to address gaps \textbf{(1)}, \textbf{(2)}, \textbf{(4)}, and \textbf{(5)}.



\section{Co-Creating LLM Solutions}
We examined AI/LLM tool adoption by co-creating tools alongside analysts and observing how they were incorporated into practice. Tools were developed in response to analysts’ expressed needs and field observations, then iteratively demonstrated during weekly or bi-weekly sessions. This process enabled analysts to guide development without adding to their workload or disrupting existing workflows. Over a six month period, we developed three security tools and one integrated security platform. 

To supply the tools with the explicit information of the SOC analysts
we relied on RAG (Retrieval-Augmented Generation)~\cite{gupta2024comprehensive,rahman2024retrieval},
a technique in LLM application development
where documents that have been turned into vector embeddings are
appended to the query prompt to add additional context to the LLM
during inference.
We use RAG and the associated
vectorstores for serving the SOC-specific data to the LLM during
response generation.

\begin{figure}[t]
  \centering
  \includegraphics[width=1\columnwidth,trim=5 5 5 5,clip]{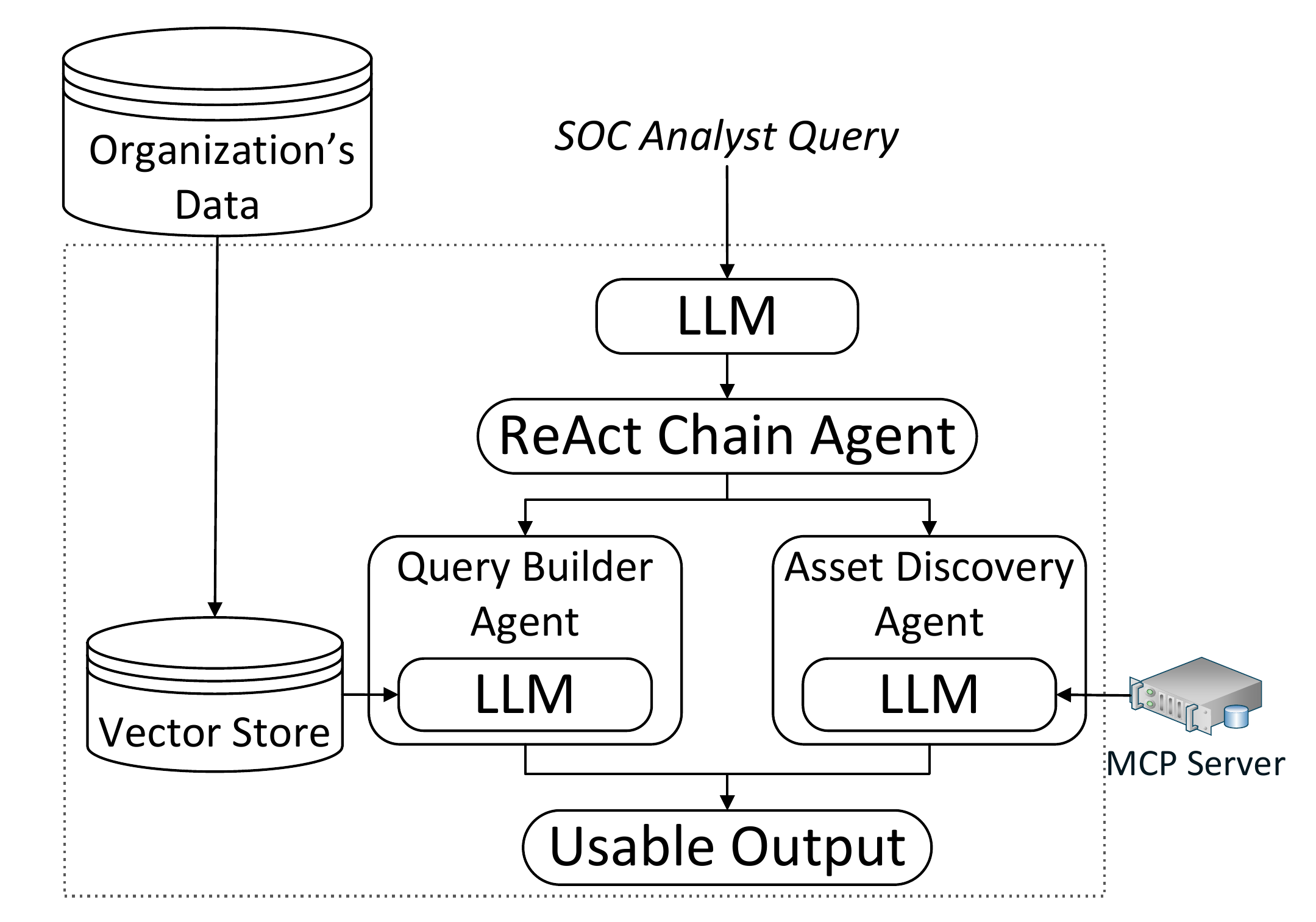}
  \vspace{-.2in}
  \caption{LLM Security Platform -- The platform receives an analyst’s query and uses an LLM with a ReAct-based agent to determine the appropriate tool or action. Selected agents interact with organizational data and vector stores via the Model Context Protocol (MCP), producing a structured, usable output for the analyst.}
  \label{figure:socplatform}
\end{figure}

This section focuses on the platform and three tools deployed within the SOC (see Figure~\ref{figure:socplatform}), highlighting how early workflow disruptions led to non-adoption and how iterative refinement, grounded in analyst feedback, ultimately supported sustained integration. 

\begin{table*}[t]
\small
\centering
\label{tab:roles_pain_points}

\begin{tabular}{
    p{0.22\linewidth}|
    p{0.24\linewidth}|
    p{0.46\linewidth}
}
\toprule
\rowcolor{gray!15}
\textbf{Role} & \textbf{Task} & \textbf{Pain Points} \\
\midrule

SOC Analyst & Root Cause Analysis & 
Tight time requirements to parse various amounts of distributed data. Can be challenging to decide on root cause when human error is involved. Can cause scheduling conflict if analysis needs to be conducted in parallel with existing tasks. \\ \midrule

SOC Analyst & Query Building &
There are of a variety of query languages for SIEMs. The SIEM used by the organizations provides example, but they are reported and observed to be limited in-scope. A query builder was provided by the security vendor for their platform, however due to the lack visbility of specific organizational resources limits the creation of readily available query statements.  \\ \midrule

Vulnerability Management & Asset Discovery &
Requires multiple highly repitive tasks to complete by analyzing distributed data sources. A highly critical and time sensitive task. Requires the use of API code to collect necessary reporting data. \\ 


\bottomrule
\end{tabular}
\caption{Operational Tasks and Pain Points -- SOC roles, recurring tasks, and workflow bottlenecks identified through embedded fieldwork, which informed the selection of initial targets for LLM-based tool development.}
\end{table*}

\subsection{The LLM Security Platform}
An initial demonstration of the security tools sparked discussion of not wanting the tools to be widely dispersed or distributed across multiple platforms or domains.
To achieve this design requirement LangChain framework was used which has a feature called the ReAct chain~\cite{yao2022react} which provided tool dispatching using LLM-based reason and action generation using supplied context for registered tools. 

Two issues were observed during the development stage: \textbf{(1)} due to the demand for GPUs supplied by our cloud service provider we were only able to obtain a 16GB slice of a NVIDIA A10 GPU, and \textbf{(2)} during initial development we observed highly variable ReAct responses between two models, Mistral:7B~\cite{jiang2023mistral} and Llama3.2:8b~\cite{meta2024llama3}. 

These observed issues led to us performing a comparative test of the LLM-platform to decide which model would perform best under our constrained environment while ensuring consistency of output using the ReAct Chain. To facilitate trust in the tools, the LLM was designed to perform correct and predictable tool dispatching while producing meaningful and usable outputs. The tests were conducted using a Macbook Pro M2 32GB and led to the use of the Mistral:7B model. Although the hardware resources were fairly modest, the needs oriented design and practical utility of the platform proved to be a decisive factor in the SOC environment. The result of the comparative tests can be found \href{https://anonymous.4open.science/r/ModelTesting}{here}.

\subsection{Model Testing}
We evaluated four models, Llama 3.2 3B (Meta), Mistral 7B (HuggingFace), Qwen3 8B (HuggingFace), and GPT-OSS 20B (OpenAI), to assess reliability, tool-calling accuracy, and response time. Five representative queries were used to test correct tool invocation and end-to-end latency. To stress-test tool selection, we introduced additional tools with similar contextual prompts. We also evaluated performance under varying RAG configurations by adjusting \textit{k\_fetch} and \textit{k\_retrieval}. 

\highlight{Tool Call Testing}
Tool-calling accuracy varied across models. GPT-OSS 20B correctly selected the intended tool in Query 1, while other models misrouted to Document Analysis. All models correctly invoked the MCP agent for Query 2, with Mistral 7B achieving the lowest response time. For Query 3, all models selected correctly, but GPT-OSS 20B exceeded the 20-minute runtime limit and was terminated. In Query 4, all models performed correctly except Llama 3.2 3B, which again misrouted. For Query 5, all models correctly utilized RAG. Notably, GPT-OSS 20B occasionally relied on pre-training knowledge instead of the RAG system. 

\highlight{Response Time Testing}
The impact of RAG parameters on performance was evaluated by varying \textit{k\_fetch} (1, 3, 5) and \textit{k\_retrieval}. Increasing both parameters improved response quality but also increased latency. For the evaluated dataset, setting both values to 3 provided a balance between accuracy and response time, producing outputs that effectively incorporated retrieved context.

\subsection{Agent Design}
There were four agents used in this work, three of the agents were explicit and one of the agents was implicit. The agents were designed to sub-divide the task, otherwise know as task decomposition~\cite{ku2024findings}, the three explicit agents we utilized were \textbf{(1)} the asset discovery agent, \textbf{(2)} the query builder agent, and \textbf{(3)} the MCP agent. The one implicit agent was \textbf{(4)} the ReAct agent which was used to decide which of the three explicit agents were to be used.


\highlight{Query Builder}
The tool was designed to assist with constructing query statements by allowing a locally run model within the organization's infrastructure that wouldn't violate confidentiality requirements as stated by the AI governance. 

The initial design of this tool was to take user prompt for a specified query statement request and using a RAG system 
which would access a vectorstore populated with examples of query statements as the knowledge base it would inform the LLM how to predict which tokens to use for generating query statements. A key feature for using RAG and vectorstores for this generation approach is that the knowledge base vectorstore would grow through repeated usage, this feature was discussed numerous times with analysts on how it would provide the analysts a growing database of shared experience. The process for vectorstore retrieval worked by taking the query prompt and retrieve the three most relevant documents to the user prompt from the vectorstore and supply it to the query. The LLM would then generate the query statement and supply it to the user. 

\highlight{Asset Discovery}
The asset discovery tool was designed to automate the task of detecting which assets on the organization network were not properly reporting to the security vendor's platform. To accomplish this, the tool was designed to accept a user query for asset discovery. It then issued an API call to the API tool registered with the MCP server in order to retrieve information about the asset’s reporting history. 

It would return back this information and then process the log data to determine the reporting status, checkpoints based on status history were created to prevent creating duplicate flaggings. The tool would then take the assets name from the API data and make another API call via the MCP server to access the datastores which contained the assets name, system information, and corresponding asset owner information. The LLM would use all of the collected information to summarize and draft an email for all assets currently marked as \textit{``not reporting''} or \textit{``stale''} and wait for for the user to respond with modifcation requirements or to send. 

\subsection{Root Cause Analysis} The RCA tool was implemented to help analysts convert heterogeneous incident artifacts (JSON/YAML objects or raw log text) into structured, SOC-ready incident narratives. To satisfy confidentiality and governance requirements, all generation is performed using locally hosted Ollama~\cite{Ollama} models so incident data remains within the organization’s infrastructure.

The RCA pipeline begins by ingesting a single incident artifact through file upload or paste mode. If the input is structured (JSON/YAML), the tool normalizes it into a consistent schema (e.g., event type, timestamp, users, IPs, host/asset identifiers, process and URLs/domains). If the input is unstructured, the system preserves the raw logs and uses them directly as the incident representation.

To ground outputs in organizational guidance, the tool supports an optional RAG workflow. Analysts can upload playbooks, policies, and prior RCA documents, which are chunked and indexed into a persistent Chroma~\cite{Chroma} vectorstore using local embeddings. For each incident, the tool constructs a retrieval query from available indicators and performs hybrid similarity search to select top-k context chunks, which are embedded into the RCA prompt with citation anchors.

The RCA generator produces a structured report containing: (1) what happened, how, and why; (2) ranked contributing factors with explicit evidence; (3) mitigations scoped by time horizon (0–24h, 1–7d, >7d); (4) data gaps and assumptions; and (5) advisory triage tags and suggested MITRE ATT\&CK~\cite{MITREATTACK} mappings. Prompt constraints require the model to reuse concrete indicators from the incident artifact and to state “Not available in provided data” rather than inferring missing fields. This was done to keep the outputs in check and avoid hallucinations. Finally, the tool logs key interaction events (input ingestion, retrieval metadata, generation outputs, and analyst feedback) to an append-only JSONL audit trail to support iterative evaluation and refinement.
\subsection{Early Integration Challenges}
The initial development of the query builder used examples for constructing query statements from the security vendors website, where the examples were obtained using a webscraper and then processed into a uniform format, within that set of examples there was one hand-crafted example by the embedded researchers based on their experience for conducting an investigation of an alert regarding excessive login failures. 

\highlight{Initial Demonstration and Use}
The initial demonstration to the SOC team used both security-vendor examples and a researcher-crafted example. After deployment, we observed that LLM queries not informed by the researcher-crafted example produced lower-quality outputs, leading to negative feedback and workflow disruption. Analysts reported similar issues with the vendor-provided query builder, indicating that the examples used to guide query construction were critical to output quality.

Another issue that was observed during the analysts initial use of the query builder tool was that it was incapable of ingesting alert specific information, this resulted in query statements that had generic or hallucinated values within the parameter fields. It was possible to provide these values through the prompt window, but it was noted that supplying log information and prompting was undiserable.

A pain point for the RCA tool was the limited access
to historical RCA reports. Formal RCA documents are rarely
produced, typically only for severe or escalated incidents. This limited access constrained our ability to ground
early prototypes in real-world reference material
or to train and validate the tool against a large corpus of
completed analyses. 

\highlight{Limited Benefits}
Development of the asset discovery tool was disruptive by a challenge in policy. As development of the asset discovery tool progressed, there was a need for the use of emerging technologies and development methods. In the case of the asset discovery tool we found after initial development that providing the information to the LLM manually, did not aid the analysts in meaningful way. The quantity and repetiveness for the asset discovery workflow in a sense demanded automation and to achieve this it required the LLM to access the data provided by the API autonomously. However, our initial proposal to the organization's AI governance team did not include accessing resources from assets beyond what was available to the server which hosted the LLM. This presented a pause in development and led to a slow down in the tools adoption while a proposal and policy was developed for autonomous internal-API calling by an LLM.

\subsection{Workflow Aligned Iterative Refinement}
Feedback from tool usage and observations during analyst shadowing informed the refinement of features. This increased analyst control over data supplied during prompting, incorporated workflow-derived query examples, and guided discussions around policies and requirements for securely enabling autonomous LLM interactions with internal APIs.

\highlight{Improvements to the Query Builder} 
The first step taken towards improving the query builder tool was adding in an additional text window for supplying logs, this fufilled the request made by the analysts to seperate where to supply logs and the prompt. This modification added the supplied text to the data pipeline for the RAG system thus using the data supplied as an additional document to retrieve from for future use. The next step was adding analyst specific example documents to the RAG vectorstore. However, getting the documents took time, and getting the examples from the analysts took more time because this was an added task to their already busy routine. These changes prompted discussions with the SOC director, during which the proposed template and its benefits for providing examples were presented. Following this, analysts were instructed to contribute workflow-derived examples. Together, these changes were associated with higher observed adoption and were reported as substantially less disruptive than earlier iterations. Analysts reported more frequent use of the tool to support their work and demonstrated increased receptivity to follow-up feedback and data collection.

\highlight{Improvements to the Asset Discovery Tool}
Improving the asset discovery tool was a different challenge, that was a mix of technical and non-technical aspects. The technical aspects were understanding and describing the security implications of deploying the MCP with access to internal API calls that allowed for observing which assets on the organization's network were properly reporting to the security vendor's platform and which assets presented a blindspot. The non-technical aspects were the planning and design of a proposal which described how we would mitigate and prevent actions that would allow the LLM and the MCP server to be used to violate security, privacy, and trust. This led to various discussions with the AI governance on how we would use traditional methods of authentication and authorization for securing the use of the LLM and MCP server. While these discussions did not lead to an additional approved proposal the analysts were still using the tool with performance issues from the GPU limitations and it being a manual process due to the lack of API support from the MCP server.

\highlight{Improvements to the RCA Tool} Development for the RCA Tool was through an iterative refine-and-test cycle.
The prototypes were deployed within the SOC, and analysts
used them on real incidents over multiple iterations. After
each trial, feedback was gathered through observations, think-aloud sessions, and debrief interviews, then rapidly incorporated changes. Each iteration brought the tools closer to the
SOC’s needs. By the second demo, the analyst’s feedback
had grown more positive, they reported that LLM suggestions
were becoming more relevant and saw clear time-saving
benefits, coupled with remaining requests like improving the
UI/UX and adding some guardrails for the LLM outputs. This participatory refinement aligned the solutions with user expectations.

\section{Findings Shaping an Extended SECI Model}
This section presents qualitative findings that inform an extension of the Tool-Oriented SECI model to an LLM-Oriented SECI framework, highlighting how LLM companions reshape knowledge conversion and tool integration in practice.

\subsection{Qualitative Findings}
We present qualitative findings from ethnographic engagement with SOC analysts, showing how operational constraints, co-creation, and trust shaped LLM tool adoption, design, and workflow-aligned use as analyst-enablers.
We use pseudonyms such as ``P4'' and ``P11'' to refer to analysts who participated in the research.

\highlight{Ethnographic Insight Reveals Operational Constraints} 
Early immersion revealed that the core challenges were not simply a lack of automation, but fragmentation across tools, teams, and objectives. P4 explained how analysts routinely moved between multiple platforms to complete a single task: using the SIEM for vulnerability data, consulting external sources for risk scores and remediation guidance, reviewing official CVE databases for documentation, and manually correlating information across these distributed systems. These disconnects between platforms and processes created cognitive strain and increased workload. 

P11, a senior security executive, commented that the fieldworkers were having \textit{``conversations that should be happening were often not.''} Analysts were focused on immediate incident response, and were not able to make long term design decisions that could improve productivity. P8 noted \textit{``any kind of automation that reduces the workload of analysts is welcome.''} These observations showed that the real need was not full automation of analyst work, but targeted support that reduced cognitive load and reclaim analyst workhours. The goal became to streamline information gathering, correlation, and summarization so analysts could focus on higher-value tasks. This ethnographic grounding helped avoid premature automation and positioned LLM integration as a tool to strengthen analysts’ capabilities, as an \textbf{\textit{analyst enabler}}.

\highlight{Co-Creation makes Tacit Knowledge Explicit} 
Interest in AI was strong across teams, with pentesting, phishing detection, incident reporting, vulnerability triage, and workflow automation identified as potential LLM use cases. Through brainstorming sessions and workflow sketching, we combined the analysts' domain knowledge with the fieldworkers' technical expertise to formulate solutions that were both useful and technically feasible. Our decision to develop the
RCA, Query Builder, and Asset Discovery tools were well
received by many of the analysts despite the initial set of
issues identified.

This process externalized tacit knowledge embedded in social interactions and informal coordination. For example, analysts emphasized the importance of reducing user communication delays during investigations, stating that \textit{``reducing that time is critical.''} Iterative refinement incorporated this feedback, resulting in features that generated richer contextual data. Early usability challenges were addressed through UI refinements, including the addition of tool tips and tutorials, enabling analysts to use the tools more effectively. Through this iterative loop, the tools adapted to real-world practice, strengthening analysts' sense of ownership and involvement. P8 remarked that they had \textit{``never seen a project where [outsiders] come in and build tools with us according to our needs,''} underscoring how co-building distinguished this approach from vendor-driven deployment.

\highlight{Trust as an enabling factor} 
Adoption depended on institutional trust along with usability. Early in the engagement, the internal advocate (IA) played a critical mediating role, lending legitimacy to the fieldworkers' presence. The importance of this role became apparent when simply referencing the IA or including them in communications was sufficient to secure approvals, drawing on the trust the IA had already established within the organization. Over time, analysts engaged directly with the fieldworkers, indicating transferred trust.

\highlight{AI enablement and workflow integration} 
P11 articulated the parameters of a successful endeavor: \textit{``successful initiatives must (1) be replicable and capable of being institutionalized, and (2) produce outputs that yield corporate value through tools that are usable in the operational environment.''} 

Analysts reported reduction in investigative time and cognitive load, particularly in consolidating distributed data sources. P5 stated, \textit{``We use these things, when we need to use it, it helps us learn more,''} reflecting companion-style usage rather than dependence. Similarly, P8 noted, \textit{``Good to get an outside perspective, tools are not the regular tools, these are some additional tools from what they get from the vendors. [Vendor] has added a query generator, this tool seems to be better than [Vendor]'s.''} 

The RCA tool was well received by the analysts with P7 emphasising \textit{``RCA tool saves a lot of time. I start my investigation with this to get an approach for the investigation, understand the root cause of the alert. I have been using it regularly. I am a new recruit and I believe this can also help in training new recruits who are learning how to start the investigations.''} The asset discovery also integrated well with the analyst workflows, \textit{"Has had two successful uses based on past usage of other users. Changed the asset device name after finding something someone else used."}. The use of RAG for query builder drew positive reception as P9 mentioned  \textit{``a benefit to the team as a whole by allowing the team to learn from each other''} by providing a growing database of shared experience. P5 reinforced this point, noting that \textit{``Everyone is using LLM to increase efficiency but the analysts cannot put alerts into ChatGPT so homegrown LLM tools like these will really help.''} Following the conclusion of the embedding, we conducted follow-up discussions to evaluate the sustained use and integration of the tools within analysts' workflows. P7 noted that the tool helped with \textit{``reducing the lag from having taken a month break for the holidays''}. P5 explained, \textit{``I don't use this tool everyday, but when I need it, I use it. It helps with things I don't know''}. These reflections indicate that the tools were adopted and retained as practical resources that supported analysts’ ongoing operational needs.

\highlight{New Ideas} 
The analysts came up with more ideas for new tools and new features in existing tools, which was another metric of success of the ethnographic co-creation approach led by trust building. Some ideas that came up were: \textit{``a tool that could tell us that this alert also came on 11 June then the analyst can go back and look how it was handled back then and it would save him a lot of time''} and \textit{``pattern recognition in the alerts, if the SOC receives a particular type of alert weekly or biweekly or if there is a pattern to that they could probably look into it. Another idea is about false positives, if there is any way to automate the process where a tool gives out a probability that there is a X\% chance this is a FP.''}

\subsection{LLM Oriented SECI}
\begin{figure}[t]
  \centering
  \includegraphics[width=1\columnwidth,trim=5 5 5 5,clip]{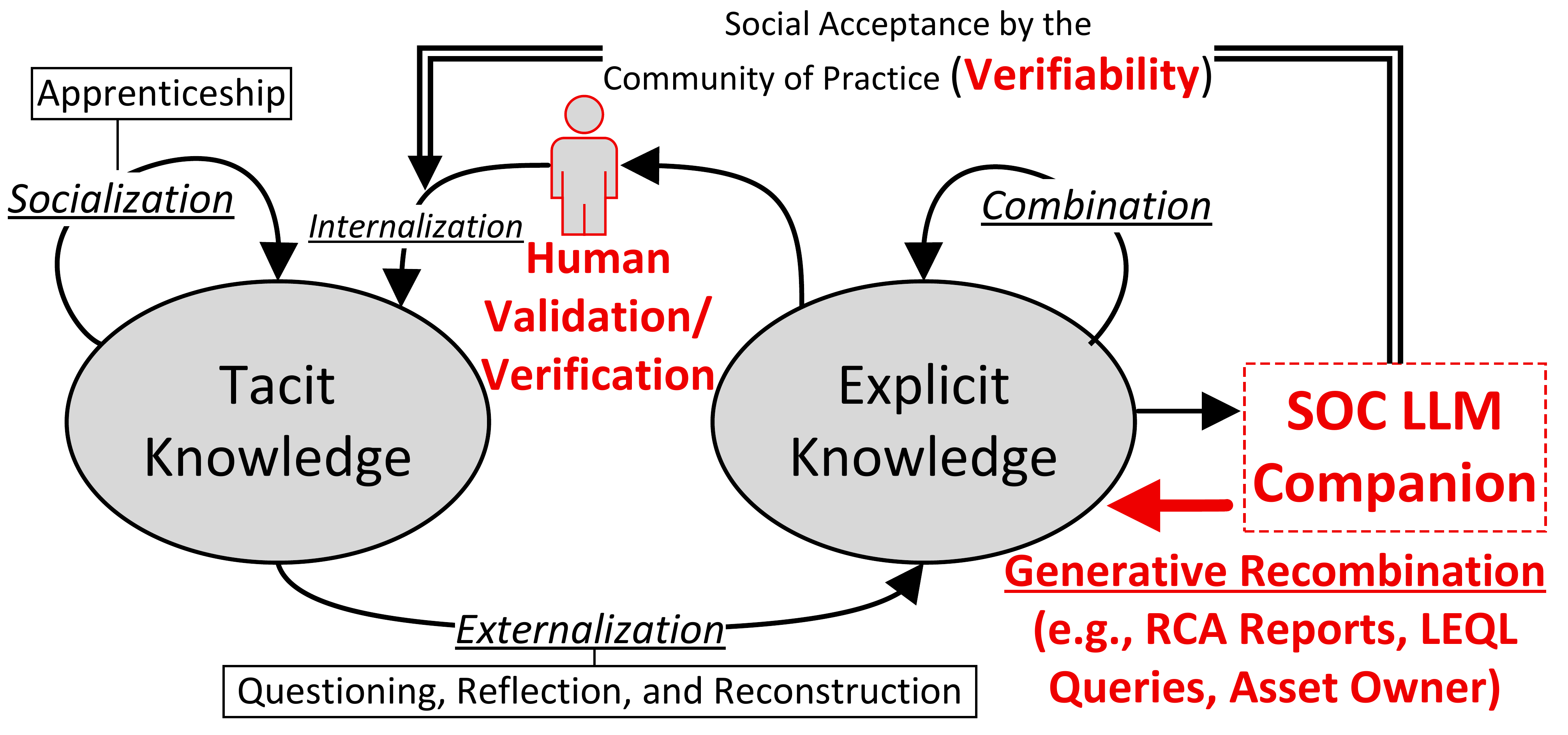}
  \vspace{-.1in}
  \caption{LLM-Oriented SECI Model -- Extension of the tool-oriented SECI framework in which the red components denote the proposed update, introducing LLM-based companion systems that support generative recombination, and human validation \& verification.}
  \label{figure:updatedtool}
\end{figure}

We observed a recurring pattern that extends tool-oriented knowledge conversion in two important ways described below. While our findings align with the general trajectory of tacit knowledge being surfaced, incorporated into tools, and integrated into practice, the introduction of LLM-based companion systems changed how explicit knowledge was produced and how it entered routine workflow as shown in Figure~\ref{figure:updatedtool}. 

 \highlight{Generative Recombination: LLMs as Source of new Explicit Knowledge}
In tool-oriented SECI~\cite{sundramurthy2014IEEE}, tools function primarily as outputs of explicit knowledge. Once tacit workflow insights are  formalized, they are embedded into algorithms, tools, or models. The tool then operationalizes this explicit knowledge. In this formulation, the tool is a stabilizing mechanism: it stores, enforces, and scales previously articulated logic. LLM-based systems extended and added a new dimension to this paradigm. We did not simply encode workflow logic into static scripts. Instead, we populated vector stores with organization-specific query examples, investigation artifacts, playbooks, prior RCA reports, and structured documentation. This enabled RAG, allowing the system to ground its outputs in locally validated explicit materials rather than relying solely on base model priors. 

However, unlike traditional tools, these LLM-based systems did not merely store or execute explicit logic. They actively generated new structured explicit artifacts. The examples, included the (i) RCA reports which synthesized alert data into structured summaries, (ii) Query builder outputs constructed from contextual prompts and retrieved examples, and (iii) Asset discovery outputs that consolidated API responses and prior documentation into analyst-ready representations. These outputs are not static rule executions. They were newly generated structured artifacts derived from recombining existing explicit materials (logs, documentation, prior examples, retrieved context, playbooks, etc). We characterize this as a \textit{generative recombination layer} within the explicit domain. Extending the SECI Combination phase (explicit$\rightarrow$ explicit), the LLM dynamically recombined stored explicit artifacts via RAG to produce candidate explicit outputs that did not previously exist in fixed form. This shifted the role of the \textit{``tool''} from a passive container of explicit knowledge to an \textit{active co-producer.} It transformed the Combination phase from a static consolidation process into a dynamic, iterative mechanism embedded directly within everyday workflow execution.

\highlight{Trust Calibrated Internalization}
However, the presence of generative outputs altered the trajectory from explicit artifact to tacit practice. In the tool-oriented SECI model, once a tool is deployed and demonstrates operational utility, its outputs are gradually incorporated into routine practice through repeated use. Explicit artifacts stabilize, become familiar, and are  internalized as part of analysts' tacit knowledge.

In our study, this progression was neither automatic nor linear. LLM-generated artifacts did not flow directly into tacit practice. Instead, they were subjected to iterative cycles of human validation, contextual verification, and selective adoption. Analysts treated model outputs as provisional hypotheses rather than authoritative results. Before integration into workflow routines, each artifact had to pass through structured and informal evaluative checks. 
Specifically, analysts assessed outputs along three dimensions. First, \textit{interpretability}: could the result be quickly inspected and mentally simulated against known ground truth or expected patterns? Second, \textit{reliability}: did the model produce stable and consistent outputs across similar inputs, or did responses vary unpredictably? Third, \textit{workflow fit}: did the artifact reduce cognitive load and investigative friction, or did it introduce additional verification overhead that offset its efficiency gains?

Only after satisfying these criteria did explicit LLM-generated artifacts transition toward internalization. Even then, internalization was conditional and partial. Analysts often embedded the outputs, as structured starting points that accelerated subsequent reasoning. In this sense, generative explicit knowledge required trust calibration before it could become tacitly embedded. The internalization phase was therefore mediated by analyst judgment, domain expertise, and situated context, rather than by repeated tool exposure alone.

All in all, our findings suggest that LLM companion systems reshape the Tool-Oriented SECI Model. Tacit knowledge is externalized through co-creation; 
operational constraints are incorporated into the tools;
the LLM generates new explicit artifacts through recombination;
analysts evaluate them; and
validated outputs are internalized into tacit knowledge.


\section{Discussion and Limitations}
We examine LLM adoption in cybersecurity operations through a sociotechnical, practitioner-centered lens. Our findings show that integration challenges extend beyond technical performance. They mostly involve how LLM-generated artifacts are validated and verified within existing workflows. Adoption emerged through co-creation, with analysts shaping tool behavior and integration rather than passively consuming outputs. By adopting tool-oriented SECI in our fieldwork, we show that LLM-based companion systems extend knowledge conversion by introducing generative recombination and trust-calibrated internalization in the transition from explicit to tacit knowledge. 

\highlight{Methodological Scope, Limitations, and Future Work}
This study is based on an embedded engagement within a single SOC, which may limit generalizability to organizations with different structures, maturity levels, or governance constraints. Tool evaluation emphasized workflow alignment, perceived usefulness, and longitudinal adoption rather than controlled performance benchmarks across diverse incident types; thus, we do not claim universal performance superiority of LLM-based tools. Additionally, because LLM capabilities and AI governance frameworks are rapidly evolving, adoption dynamics may shift over time. 
Future research could include multi-site longitudinal studies across diverse SOC environments to examine how organizational maturity, governance structures, and risk tolerance shape trust calibration and co-creative adoption dynamics. Comparative analysis can help distinguish between context-dependent settings and aspects generalizable across settings of LLM Tool-Oriented SECI.

\highlight{A Sociotechnical Approach to Tool Building} 
By performing an ethnographic study and co-creating tools alongside cybersecurity professionals, the fieldworkers were able to build tools that the analysts both understood and trusted. The fieldworkers were able to engage with the analysts frequently and perform the tasks of their job to get unique depth of understanding how they perform their work and understand their workflows. The act of socializing in a zone of proximal development enabled us to gather highly detailed tacit knowledge made explicit which was used to inform the LLM models specifically on the use-cases necessary for their operational environment. LLM adoption in SOCs is difficult because generative outputs are probabilistic and sometimes inconsistent. This amplifies existing concerns about trust and reliability. Our findings indicate that sociotechnical co-creation enhanced the adoption pathway of LLM in security operations. 

Importantly, LLM-based tools were not deployed as static solutions but they evolved through iterative refinement driven by practitioner feedback. Analysts actively shaped when, and how LLM outputs were generated and accepted, shifting adoption from top-down vendor driven approach towards collaborative knowledge conversion. The SECI framework guided this LLM integration in ways that preserved analyst judgment while enabling effective augmentation.

\highlight{LLMs extend the SECI Model} 
While LLM-based companion tools generated queries, RCA drafts, and remediation messages, these artifacts did not automatically become part of SOC practice. Instead, they were incorporated only after processes of validation and governance negotiation.

In tool-oriented SECI, internalization (explicit$\rightarrow$tacit) occurs as explicit artifacts are absorbed into routine work through use. Our findings suggest that in AI-augmented environments this transition is mediated rather than direct. Because LLM outputs are probabilistic and dynamically generated, they are treated as provisional. Before becoming tacit practice, they must be validated and verified for interpretability, reliability, and workflow fit. 

\highlight{Human-AI Knowledge Loop} 
Accoring to the tool-oriented SECI model, tools are the final outcome of explicit knowledge. In our setting, LLM companions acted as generative participants in explicit knowledge production: through RAG, stored examples and documentation were recombined into new artifacts. This introduces a hybrid human–AI knowledge loop: Tacit expertise surfaces through collaboration; workflow expectations are operationalized within LLM-based systems (e.g., vector store and RAG); the tool generates explicit artifacts; analysts validate and refine them;
and ultimately validated outputs become embedded into tacit workflow.

All in all, this indicates that LLM systems should not be evaluated solely as automation tools. They can function as dynamic collaborators in explicit knowledge production. However, internalization remains human-centered and institutionally constrained. This explains both early skepticism and eventual selective stabilization observed in our fieldwork.

\section{Conclusion}
This study examined the integration of LLM-based companion systems within a security operations center through a sociotechnical, practitioner-centered approach. Adoption emerged through co-creation: analysts actively shaped tool behavior, constraints, and deployment boundaries, ensuring alignment with operational requirements. By leveraging and extending Tool-Oriented SECI, we demonstrate that LLM systems introduce two critical dynamics into knowledge conversion: generative recombination and trust-calibrated internalization. Our extension to LLM Tool-Oriented SECI highlights that in generative AI contexts, tools not only incorporate explicit knowledge but also actively produce new explicit artifacts that reshape how tacit expertise develops over time.


\section{Ethics Statement}
This study involved security analysts and managerial staff and was conducted in accordance with established ethical standards. Research activities took place between July and November 2025 and were approved by the relevant Institutional Review Boards (IRBs). Written consent was waived; participants provided informed oral consent and received IRB-approved documentation describing the study, procedures, and their right to withdraw.

\highlight{LLM Usage Considerations} LLMs were used solely for editorial purposes, and all outputs were reviewed by the authors for accuracy and originality.
\newpage


\section{Acknowledgments}
This work is supported by the National Science Foundation under awards no. 2143393, no. 2235102, 
and the Office of Naval Research under award no. N00014-23-1-2538. Any opinions, findings and conclusions or 
recommendations expressed in this material are those of the authors and do not necessarily
reflect the views of these agencies. 

We are grateful to our organizational collaborators for welcoming us into their environments and for their partnership in enabling this work through shared learning and co-creation.

\bibliographystyle{IEEEtran}
\bibliography{References,soc_ai}
\appendix
\twocolumn
\section{Codebook}
\label{appendix:codebook}
\noindent
\begin{strip}
\centering
\captionof{table}{Thematic Structure: High-Level Themes, Definitions, and Associated Codes}
\label{tab:themes_codebook_appendix}
\small
\begin{tabularx}{\textwidth}{p{0.23\textwidth} p{0.37\textwidth} p{0.35\textwidth}}
\toprule
\textbf{Theme} & \textbf{Definition} & \textbf{Codes} \\
\midrule

\textbf{Ethnographic Insight} &
Embedded anthropological approach actively reexamined operational pain points, surfaced tacit constraints, and supported the co-creation of LLM tools that were successfully adopted in practice. &
anthropology; academics; interdisciplinary; research questions; knowledge transfer; gap between academic and industry people; fieldworker's research background; internal advocate; CISO; soc director; upper management; long term goals of org; objectives of org; organisation. \\

\addlinespace

\textbf{Operational and Structural Constraints} &
Persistent workflow bottlenecks, expertise gaps, and infrastructural limitations created cognitive and temporal strain within security operations, generating the need for augmentation rather than automation replacement. &
challenges; lack of time and expertise for tool building; different roles. different expertise.; task allocation; pentesting; rapid7; RCA; table top exercises; side projects; critical thinking; familiarise with organisation; problem identification; security tool gap; distributed data; disparate workflows; cognitive friction; infrastructure latency; communication latency ; environmental mismatch \\

\addlinespace

\textbf{Co-Creation as Tacit Knowledge Externalization} &
Through participant observation, shadowing, demos, and iterative refinement, tacit operational expertise was made explicit and embedded into tool design, transforming socially situated knowledge into structured technical artifacts. &
participant observation; fieldwork; talking to people; tacit knowledge; requirement gathering; idea formulation; demo; feedback on project; feedback on tools; tool building; solution; collaboration; socialization; discussing workflow; human non-intelligible information; data confidence; work enablement; co-building; reclaiming work hours \\

\addlinespace

\textbf{Governance, Control, and Human Authority} &
AI integration was negotiated through governance constraints, security assurances, and preservation of analyst authority. Adoption depended on bounded autonomy, human-in-the-loop positioning, and institutional trust-building. &
concerns with LLM; human in the loop; trust building; self manageable tools; automation; analyst enabler;  
concerns with LLM; human in the loop; trust building; self manageable tools; automation; analyst enabler. \\

\addlinespace

\textbf{Situated AI Enablement and Workflow Integration} &
Adoption emerged when LLM systems aligned with existing workflows, reclaimed work hours, and demonstrated utility without disrupting operational practice. Tools were embraced when positioned as enablers embedded within routine work. &
AI; AI adoption; LLM; tool; tool usage; positive reception; solution. \\

\bottomrule
\end{tabularx}
\end{strip}
\normalsize

\end{document}